\documentclass[%
 aip,
 jmp,%
 amsmath,amssymb,
 reprint,%
]{revtex4-1}

\usepackage{graphicx}% Include figure files
\usepackage{dcolumn}% Align table columns on decimal point
\usepackage{bm}% bold math
\begin{document}

\preprint{AIP/123-QED}

\title{Mean first-passage time for random walks in general graphs with a deep trap}
% Force line breaks with \\
%\thanks{Footnote to title of article.}

\author{Yuan Lin}
\author{Alafate Julaiti}

\author{Zhongzhi Zhang}
\email{zhangzz@fudan.edu.cn}

\affiliation {School of Computer Science, Fudan University,
Shanghai 200433, China}

\affiliation {Shanghai Key Lab of Intelligent Information
Processing, Fudan University, Shanghai 200433, China}

%\affiliation {State Key Laboratory for Novel Software Technology, Nanjing University, China}

\date{\today}% It is always \today, today,
             %  but any date may be explicitly specified

\begin{abstract}
We provide an explicit formula for the global mean first-passage time (GMFPT) for random walks in a general graph with a perfect trap fixed at an arbitrary node, where GMFPT is the average of mean first-passage time to the trap over all starting nodes in the whole graph. The formula is expressed in terms of eigenvalues and eigenvectors of Laplacian matrix for the graph. We then use the formula to deduce a tight lower bound for the GMFPT in terms of only the numbers of nodes and edges, as well as the degree of the trap, which can be achieved in both complete graphs and star graphs. We show that for a large sparse graph the leading scaling for this lower bound is proportional to the system size and the reciprocal of the degree for the trap node. Particularly, we demonstrate that for a scale-free graph of size $N$ with a degree distribution $P(d)\sim d^{-\gamma}$ characterized by $\gamma$, when the trap is placed on a most connected node, the dominating scaling of the lower bound becomes $N^{1-1/\gamma}$, which can be reached in some scale-free graphs. Finally, we prove that the leading behavior of upper bounds for GMFPT on any graph is at most $N^{3}$ that can be reached in the bar-bell graphs. This work provides a comprehensive understanding of previous results about trapping in various special graphs with a trap located at a specific location.
\end{abstract}

% PACS, the Physics and Astronomy
                             % Classification Scheme.
%\keywords{Suggested keywords}%Use showkeys class option if keyword
                              %display desired
%05.40.Fb Random walks and Levy flights
%61.43.Hv  Fractals; macroscopic aggregates (including diffusion-limited aggregates)
%05.45.Df Fractals
%05.60.Cd Classical transport
%05.40.-a Fluctuation phenomena, random processes, noise, and Brownian motion
%89.75.Hc Networks and genealogical trees

\maketitle

%\tableofcontents

\section{Introduction}

Trapping problem first introduced in the
seminal work of Montroll in~\cite{Mo69} is a kind of random walk in which a perfect trap is located at a given position, absorbing all particles (walkers) visiting it. Trapping process is closely related to various other dynamical processes in diverse complex systems, including lighting harvesting in antenna systems~\cite{BaKlKo97,BaKl98}, energy or exciton transport in polymer systems~\cite{SoMaBl97}, charge transport in amorphous solids~\cite{No97}, phtoton-harvesting in cells~\cite{WhGo99}, and so forth. In view of its practical significance, it is of great interest to study trapping problem in a large variety of complex systems.

The highly desirable quantity of trapping problem is mean first-passage time (MFPT)~\cite{Re01}. The MFPT for a node $i$ to the trap is the expected time taken by a particle starting from $i$ to first arrive at the trap. The average of MFPT over all source nodes in the system other than the trap is often called global mean first-passage time (GMFPT), which is very helpful for understanding the kinetics of the whole trapping process and provides a useful indicator for the trapping efficiency. It is thus very important to evaluate GMFPT on different systems. Thus far, there has been considerable interest in computing GMFPT on various systems in order to uncover the effect of system structure on the behavior of GMFPT, such as regular lattices with different dimensions~\cite{Mo69,GLKo05,GLLiYoEvKo06}, the Sierpinski fractals~\cite{KaBa02PRE,KaBa02IJBC, BeTuKo10}, the $T-$fractal~\cite{KaRe89,Ag08,HaRo08,LiWuZh10,ZhWuCh11}, Cayley trees~\cite{WuLiZhCh12} and Vicsek fractals~\cite{WuLiZhCh12} as models of polymer networks~\cite{GuBl05,CaCh97,ChCa99,BiKaBl01,MuBiBl06,BlJuKoFe03,BlFeJuKo04}, as well as many
scale-free graphs~\cite{KiCaHaAr08,ZhQiZhXiGu09,ZhGuXiQiZh09,ZhXiZhLiGu09,AgBu09,ZhLiGoZhGuLi09,TeBeVo09,AgBuMa10,ZhYaGa11,ZhYaLi12,MeAgBeRo12}.

Although previous works have uncovered some nontrivial effects of different structural properties (e.g., scale-free behavior~\cite{BaAl99}, modularity~\cite{RaSoMoOlBa02}, and fractality~\cite{SoHaMa05}) on GMFPT, they focused on special graphs with the trap fixed on a special node. So far, the determination of GMFPT on general graphs with a trap at an arbitrary position is still open, and no quantitative formula has been obtained for GMFPT on a general graph. On the other hand, different topological features are often dependent on one another, e.g., in real systems fractal characteristic is frequently accompanied by a modular structure~\cite{SoHaMa06}. It is thus difficult to discern the roles of different structural properties on GMFPT. Sometimes, one may confuse and even distort the roles of different topologies.

In this paper, we perform an in-depth research about trapping problem on a general graph with the deep trap placed at an arbitrary node. We derive an implicit expression for GMFPT, in terms of eigenvalues and their corresponding eigenvectors of Lapacian matrix associated with the graph. On the basis of the obtained formula, we then provide a lower bound of GMFPT, and show that it can be achieved in two graphs, complete graphs and star graphs. In addition, for large sparse systems we find that the leading term of the lower bound is proportional to the system size and reciprocal of the degree of the trap node, irrespective of any individual structure. Particularly, we show that in a scale-free graph with $N$ nodes obeying degree distribution $P(d) \sim d^{-\gamma}$, where the exponent $\gamma$ characterizes the extent of inhomogeneity, the dominating scaling of the lower bound of GMFPT to a most-connected node is $N^{1-1/\gamma}$. Also, by making a comprehensive analysis of previous studies, we show that our work can unify existing results about GMFPT obtained for various scale-free graphs. Finally, we show that the GMFPT of any graph is at most $N^{3}$.

\section{Global mean first-passage time on general graphs}

In this section we study trapping problem on a general graph $G$ with $N$ nodes and $E$ edges, which is a particular random walk performed on the graph with a single trap located at an arbitrary node. The random walk considered here is a simple discrete-time random walk~\cite{NoRi04,CoBeTeVoKl07}. At each time step, the walker jumps from its current position to any of its neighbors with the same probability. Let $T_{ij}$ denote the MFPT from node $i$ to node $j$. For convenience, let $j$ be the trap node, and let $T_{j}$ denote the GMFPT to node $j$. Then, by definition $T_{j}$, the most important quantity for the trapping problem, is given by
\begin{equation}\label{T00}
T_j =\frac{1}{N-1}\sum_{j}T_{ij}\,.
\end{equation}

In the sequel, we will use the connection~\cite{Te91} between resistance distance and MFPT to derive an explicit formula for $T_{j}$, based on which we will provide a lower bound for $T_{j}$, as well as the leading scaling for this lower bound when the graph is sparse and very large. Moreover, we will give an upper bound for $T_{j}$.

\subsection{Explicit expression for global mean first-passage time }

Let $\overline{G}$ represent the corresponding resistor network~\cite{DoSn84} of graph $G$, which is obtain from $G$ by arranging a unit resistor along every edge of $G$. Let $R_{ab}$ be the effective resistance, i.e., resistance distance~\cite{KlRa93}, between any pair of nodes $a$ and $b$ in the resistor network, which is defined as the voltage when a unit current enters one node and leaves the other. Then, based on the established relation governing MFPT and effective resistance~\cite{Te91}, we have
\begin{equation}\label{T01}
T_{ij}=\frac{1}{2}\sum_{z}d_{z}\left(R_{ij}+R_{jz}-R_{iz}\right),
\end{equation}
where $d_{z}$ is the degree of node $z$ in graph $G$.

Thus, in order to determine $T_{ij}$, one can alternatively evaluate the terms about effective resistance on the right-hand side of Eq.~(\ref{T01}). It is known that effective resistance between any two nodes can be determined by eigenvalues and eigenvectors of Laplacian spectrum of the original graph. For graph $G$ with size $N$, its Laplacian matrix ${\bf L}$ is an $N\times N$ matrix, whose elements $l_{ij}$ are defined as follows: the non-diagonal entry $l_{ij}=-1$, if nodes $i$ and $j$ are connected by a link, otherwise $l_{ij}=0$; while the diagonal entry $l_{ii}=d_i$. Let $\lambda_1$, $\lambda_2$, $\lambda_3$, $\cdots$, $\lambda_N$ be the $N$ eigenvalues of ${\bf L}$, rearranged in an increasing order as  $0=\lambda_1<\lambda_2\leq \cdots \leq \lambda_N$, and let $\mu_1,\mu_2,\cdots\mu_N$ be the corresponding mutually orthogonal eigenvectors of unit length, where $\mu_i=(\mu_{i1}, \mu_{i2}, \cdots \mu_{iN})^{\top}$. Then the effective resistance $R_{ij}$ is given by~\cite{Wu04}
\begin{equation}\label{T02}
R_{ij}=\sum_{k=2}^{N}\frac{1}{\lambda_k}(\mu_{ki}-\mu_{kj})^2.
\end{equation}
Inserting Eqs.~(\ref{T01}) and~(\ref{T02}) into Eq.~(\ref{T00}), we have
\begin{eqnarray}\label{T03}
T_{j}&=&\frac{1}{N-1}\sum_{i}\frac{1}{2}\sum_{z}d_{z}\sum_{k=2}^{N}\frac{1}{\lambda_{k}}\big[(\mu_{ki}-\mu_{kj})^2\nonumber\\
&\quad&+(\mu_{kj}-\mu_{kz})^2-(\mu_{ki}-\mu_{kz})^2\big].
%&=&\frac{1}{N-1}\sum_{k=2}^{N}\frac{1}{\lambda_k}\sum_{i}\sum_{z}d_{z}\big[\mu_{kj}^2-\mu_{ki}\mu_{kj}\nonumber\\
%&\quad& -\mu_{kj}\mu_{kz}+\mu_{ki}\mu_{kz}\big]\,.
\end{eqnarray}

Equation~(\ref{T03}) can be simplified by using the properties for related quantities of matrix $\rm{\bf L}$ listed below. According to the spectral graph theory~\cite{Ch97}, we have:
%\begin{equation}\label{T04}
%{\rm\bf L}={\rm\bf U}{\rm diag}\left[{\lambda_1,\lambda_2,\cdots,\lambda_N}\right]{\rm\bf %U}^{\top},
%\end{equation}
%where ${\rm \bf U}=[\mu_1,\mu_2,\cdots,\mu_N]$ is an orthogonal matrix obeying ${\rm\bf U} {\rm\bf U}^{\top}={\rm\bf U}^{\top}{\rm\bf U}=\rm{\bf I}$, in which $\rm{\bf I}$ is the identity matrix with order $N \times N$. In other words, the eigenvectors of matrix $\rm{\bf L}$ satisfy

(i) the eigenvectors of matrix $\rm{\bf L}$ satisfy
\begin{equation}\label{T05}
\mu_i^{\top}\mu_j=\delta_{ij}\,,
\end{equation}
where $\delta_{ij}$ is the Kronecker delta function defined as: $\delta_{ij}=1$ if $i$ is equal to $j$, and $\delta_{ij}$=0 otherwise.

(ii) the entries of eigenvector $\mu_{1}$ corresponding to $\lambda_1=0$ are all equal to each other, namely $\mu_{1i}=\sqrt{N}/N$ holds for $1\leq i\leq N$. Then, for $1 < k \leq N$,
\begin{equation}\label{AT00}
\sum_i \mu_{ki} =0.
\end{equation}

(iii) the entry $l_{ij}$ of matrix $\rm \bf L$ has the following spectral representation:
\begin{equation}\label{T06}
l_{ij}=\sum_{k=1}^{N}\lambda_{k}\mu_{ki}\mu_{kj}\,,
\end{equation}
which means
\begin{equation}\label{T07}
l_{ii}=d_{i}=\sum_{k=1}^{N}\lambda_{k}\mu_{ki}^2.
\end{equation}

The above introduced properties of matrix $\rm \bf L$ are very useful for the following derivation. We first use them to simplify  Eq.~(\ref{T03}) that can be recast as
\begin{eqnarray}\label{T09}
\langle T_j \rangle&=&\frac{1}{N-1}\sum_{k=2}^{N}\frac{1}{\lambda_k}\bigg[\sum_i\sum_z d_z \mu_{kj}^2 - \sum_i\sum_z d_z \mu_{ki}\mu_{kj}\nonumber\\
&\quad&-\sum_i\sum_z d_z\mu_{kj}\mu_{kz}+\sum_i\sum_z d_z\mu_{ki}\mu_{kz}\bigg].
\end{eqnarray}
From the above intermediary results, the four terms in the square brackets of Eq.~(\ref{T09}) can be sequentially calculated as follows:
\begin{equation}\label{T10}
\sum_i\sum_z d_z\mu_{kj}^2 = N\times2E\times\mu_{kj}^2,
\end{equation}
\begin{equation}\label{T11}
\sum_i\sum_z d_z\mu_{ki}\mu_{kj}=2E\times\mu_{kj}\sum_i\mu_{ki}=0,
\end{equation}
\begin{equation}\label{T12}
\sum_i\sum_z d_z\mu_{kj}\mu_{kz}=N\times\mu_{kj}\sum_z d_z\mu_{kz},
\end{equation}
and
\begin{equation}\label{T13}
\sum_i\sum_z d_z\mu_{ki}\mu_{kz}=\sum_i \mu_{ki}\sum_z d_z\mu_{kz}=0,
\end{equation}
where the facts $E=\sum_z d_z / 2$ and $N=\sum_i 1$ are used. Plugging Eqs.~(\ref{T10}-\ref{T13}) into Eq.~(\ref{T09}), we arrive at the explicit expression for GMFPT on a general graph with the immobile trap fixed at an arbitrary node $j$, given by
\begin{equation}\label{T14}
T_j=\frac{N}{N-1}\sum_{k=2}^{N}\frac{1}{\lambda_k}\left(2E\times\mu_{kj}^2-\mu_{kj}\sum_{z}d_z\mu_{kz}\right)\,,
\end{equation}
which is a main result of this work. It is not obvious that Eq.~(\ref{T14}) depends on any individual structural feature.
%Equation~(\ref{T14})

\subsection{Lower bound for global mean first-passage time}

After obtaining the exact formula for GMFPT, we next apply it to derive a lower bound for $T_j$.
By Cauchy's inequality, one has
\begin{small}
\begin{widetext}
\begin{eqnarray}\label{T15}
\left[\sum_{k=2}^{N}\frac{1}{\lambda_k}\left(2E\times\mu_{kj}^2-\mu_{kj}\sum_{z}d_z\mu_{kz}\right)\right]
\left[\sum_{k=2}^{N}\lambda_k\left(2E\times\mu_{kj}^2-\mu_{kj}\sum_{z}d_z\mu_{kz}\right)\right]\geq
\left[\sum_{k=2}^{N}\left(2E\times\mu_{kj}^2-\mu_{kj}\sum_{z}d_z\mu_{kz}\right)\right]^2.
\end{eqnarray}
\end{widetext}
\end{small}
Considering Eqs.~(\ref{T06}) and~(\ref{T07}) and $\lambda_1=0$, we have
\begin{eqnarray}\label{T16}
&\quad&\sum_{k=2}^{N}\lambda_k\left(2E\times\mu_{kj}^2-\mu_{kj}\sum_{z}d_z\mu_{kz}\right)\nonumber\\
&=&2E\sum_{k}\lambda_{k}\mu_{kj}^2-\sum_z d_z\sum_k\lambda_k\mu_{kj}\mu_{kz}\nonumber\\
&=&2E\, d_{j}-\sum_z d_z l_{jz}.
\end{eqnarray}
In addition,
\begin{small}
\begin{eqnarray}\label{T17}
&\quad&\sum_{k=2}^{N}\left(2E\times\mu_{kj}^2-\mu_{kj}\sum_{z}d_z\mu_{kz}\right)\nonumber\\
&=&2E\sum_{k}\mu_{kj}^2-2E\mu_{1j}^2-\sum_z d_z\sum_{k}\mu_{kj}\mu_{kz}+\mu_{1j}\sum_z d_z\mu_{1z}.\nonumber\\
%&=&2E-2E\times\frac{1}{N}-d_j+2E\times\frac{1}{N}=2E-d_j.
\end{eqnarray}
\end{small}
Equation~(\ref{T05}) shows that
\begin{equation}\label{Mat06}
\sum_{k=1}^{N}\mu_{k j}\mu_{kz}=
\begin{cases}
1, &{\rm if} \quad j=z,\\
0, &{\rm otherwise},
\end{cases}
\end{equation}
which leads to
\begin{equation}\label{Mat07}
\sum_z d_z\sum_{k}\mu_{kj}\mu_{kz}=d_j
\end{equation}
and
\begin{equation}\label{Mat08}
\mu_{1j}\sum_z d_z\mu_{1z}=\frac{2E}{N}.
\end{equation}
Therefore, Eq.~(\ref{T17}) can be simplified to
\begin{eqnarray}\label{T18}
&\quad&\sum_{k=2}^{N}\left(2E\times\mu_{kj}^2-\mu_{kj}\sum_{z}d_z\mu_{kz}\right)\nonumber\\
&=&2E-2E\times\frac{1}{N}-d_j+2E\times\frac{1}{N}=2E-d_j.
\end{eqnarray}
Combining the above obtained expressions, we have
\begin{eqnarray}\label{T19}
T_j \geq \frac{N}{N-1}\frac{(2E-d_j)^2}{2E\times d_j - \sum_z d_z l_{jz}}\,,
\end{eqnarray}
where $l_{jz}$ is the entry of Laplacian matrix $\rm \bf L$ as defined before.

Let $a_k=\frac{1}{\lambda_k}(2E\mu_{kj}^2-\mu_{kj}\sum_{z}d_z\mu_{kz})$ and $b_k=\lambda_k(2E\mu_{kj}^2-\mu_{kj}\sum_{z}d_z\mu_{kz})$. In Eq.~(\ref{T19}), the equality holds if and only if $a_2/b_2=a_3/b_3=\cdots = a_N/b_N$. We distinguish two cases. If $2E\mu_{kj}^2-\mu_{kj}\sum_{z}d_z\mu_{kz}\neq 0$ for all $2\leq k\leq N$, then the equality of Eq.~(\ref{T19}) holds if and only if $\lambda_2=\lambda_3=\cdots = \lambda_N$. In this case, only when $G$ is isomorphic to the complete graph, the equality holds~\cite{Zh08}. For the other case that there exists some $k$ such that $2E\mu_{kj}^2-\mu_{kj}\sum_{z}d_z\mu_{kz}=0$, then the equality holds if for those $k$ with   $2E\mu_{kj}^2-\mu_{kj}\sum_{z}d_z\mu_{kz}\neq 0$, the corresponding eigenvalues $\lambda_k$ are equal to each other. For example, the equality holds for the star graph, since for the star graph the term $2E\mu_{kj}^2-\mu_{kj}\sum_{z}d_z\mu_{kz}$ in Eq.~(\ref{T15}) is non-zero only when $k=N$, while for other $k$ ($k=2,3, \cdots,N-1$), $2E\mu_{kj}^2-\mu_{kj}\sum_{z}d_z\mu_{kz}$ is exactly zero.

It is easy to prove that the term $\sum_z d_z l_{jz}$ reaches its minimal value when node $j$ is linked to all other nodes in the graph excluding node $j$ itself, that is,
\begin{equation}\label{T20}
\sum_z d_z l_{jz} \geq d_j^2 - (2E - d_j)\,.
\end{equation}
Therefore,
\begin{equation}\label{T21}
T_j \geq \frac{N}{N-1}\frac{(2E-d_j)^2}{2E\times d_j - d_j^2 + 2E - d_j} = \frac{N}{N-1}\frac{2E-d_j}{d_j+1}.
\end{equation}
In this way, we have obtained a lower bound for GMFPT in terms of the number of nodes, the number of edges, and the degree of trap node, but independent of other structural parameters.

The lower bound for GMFPT given in Eq.~(\ref{T21}) is sharp since it can be achieved in some graphs. For example, it has been reported that for the complete graph with $N$ nodes, the MFPT between any node pair $i$ and $j$ is exactly $N-1$~\cite{Bobe05}. Thus, for trapping on the complete graph with the single trap positioned at an arbitrary node, the GMFPT is $N-1$, which is equal to the lower bound provided by Eq.~(\ref{T21}). Again for instance, it can be proved that for trapping on the star graphs with the trap being placed on the center node, Eq.~(\ref{T21}) can also be reached.

The obtained lower bound for GMFPT, i.e., Eq.~(\ref{T21}), provides important and useful information of a graph since it establishes a range of GMFPT in terms of the simple graph parameters. Moreover, although the lower bound can only be achieved in some very few cases, as we will see that its leading scaling can reconcile previous results on GMFPT for random walks on different networks, especially on scale-free networks.

Extensive empirical works have shown that most real networked systems are sparse in the sense that their average node degree $\langle d \rangle=\frac{E}{N}$ is a small constant~\cite{AlBa02,Ne03,BoLaMoChHw06}. Equation~(\ref{T21}) shows that in sparse graphs, when trap node is placed on a node with degree $d$, the lower bound becomes
\begin{equation}\label{T22}
T_d  \geq \frac{N}{N-1}\frac{N\langle d\rangle - d}{d+1}\,.
\end{equation}
In the limit of large network size $N$, the leading term of the lower bound is
\begin{equation}\label{T222}
T_d  \geq \frac{N \langle d\rangle}{d}\,,
\end{equation}
which is proportional to the system $N$ and the inverse degree of the trap node. We note that this dominating term has been previously reported in~\cite{TeBeVo09}, deduced by another technique, but the definition of GMFPT given in~\cite{TeBeVo09} is different from the one here that is more frequently adopted in the literature.

In addition to the sparseness, most real-life networks also display scale-free properties~\cite{BaAl99} with their degree distributions $P(d)$ obeying a power-law form $P(d)\sim d^{-\gamma}$ characterized by the exponent $\gamma$, which suggests that there exist large-degree nodes (hubs) in these networks whose degree $d_{\rm max}$ satisfying $d_{\rm max}\sim N^{1/(\gamma-1)}$~\cite{CoErAvHa01,DoGoMe08}. These nodes have a strong effect on dynamical processes occurring on graphs~\cite{Ne10}, including the trapping process discussed here. For example, when a trap is located at a hub in the pseudofractal web~\cite{ZhQiZhXiGu09}, the GMFPT exhibits a sublinear scaling with system size $N$ as $N^{\frac{\ln2}{\ln3}}$. Such sublinear scaling underlies a high trapping efficiency which was evidenced also in~\cite{ZhGuXiQiZh09,AgBu09,ZhLiGoZhGuLi09}. Equation~(\ref{T222}) implies that when a trap is fixed at a hub with degree $d_{\rm max}$ on a scale-free graph, the scaling of the lower bound for GMFPT is $\frac{N\langle d \rangle}{d_{\rm max}}$, behaving as $N^{1-1/(\gamma-1)}$ in large systems. This minimal scaling is the possible highest trapping efficiency that can be achieved in some hierarchical~\cite{AgBu09,AgBuMa10,MeAgBeRo12} and modular~\cite{ZhLiGoZhGuLi09,ZhYaLi12} scale-free graphs. % Any other future work exploring scale-free graphs with higher trapping efficiency is unnecessary.
This implies that, in the future, looking for special topologies exhibiting an efficiency higher and higher may be a kind of obsolete problem. On the other hand, besides the sublinear scaling, the GMFPT on scale-free graphs can also display linear~\cite{ZhZhXiChLiGu09,ZhGaXi10} even superlinear~\cite{ZhXiZhGaGu09,ZhLiMa11} scaling with the system size, all of which are encompassed in the framework presented here.

\subsection{Scaling of upper bounds for global mean first-passage time}

We proceed to provide an upper bound for the GMFPT on a general graph. For this case, it is difficult to provide an exact expression as that of the lower bound, but we can give a scaling of upper bound. To this end, we recall another quantity, i.e., cover time, for random walks. The cover time for a node $i$ denoted by $C_i$ is the expected time needed for the walker starting from $i$ to first visit all the nodes in the graph. % and the cover time of the graph is the maximal value of $C_i$
It has been established~\cite{Fe95} that for an arbitrary graph, the upper bound of cover time $C_i$ grows with the system size $N$ as $N^3$. Note that in any graph, it is evident that $T_{ij}\leq C_i$, implying that $N^3$ is also an upper bound for GMFPT of the graph. It is easy to prove~\cite{BrWi90} that this upper bound can be reached in the bar-bell graph of $N$ nodes, which consists of two cliques each of size $N/3$, connected by a path of length $N/3$.

\section{Conclusions}

We have studied the trapping problem on a general graph, which is a special random walk with a trap located at a certain node. Using the spectral graph theory, we have obtained an explicit solution to GMFPT, which is expressed in terms of the eigenvalues and eigenvectors of Laplacian matrix of the graph. Based on this result we have further provided a lower bound for GMFPT that can be achieved in complete graphs and star graphs. The obtained lower bound implies that in large sparse systems its leading term grows proportionally with the system size and the reciprocal of the degree of the target node. Particularly, we have demonstrated that for trapping on a scale-free graph with the immobile trap being positioned on a hub node, the minimal scaling of the lower bound can be expressed by the exponent degree distribution $\gamma$ characterizing the inhomogeneity of the graph, which can be reached in some special scale-free graphs. At last, we have deduced an upper bound for the GMFPT. Our work provides a broader view of previous researches for trapping on diverse graphs and sheds light on some aspects related to the trapping problem, e.g., lighting harvesting.

\begin{acknowledgments}
This work was supported by the National Natural Science Foundation
of China under Grant Nos. 61074119 and 11275049.
% and the Hong Kong Research Grants Council under the General Research Funds Grant CityU 1114/11E.
\end{acknowledgments}

\nocite{*}
%\bibliography{aipsamp}% Produces the bibliography via BibTeX.

\end{document}